\documentclass[prd,twocolumn]{revtex4}
\usepackage{graphicx, epsfig}
\usepackage{color}
\usepackage{subfigure}
\usepackage{mathrsfs}
\usepackage{amsmath}
\usepackage{latexsym}
\usepackage{amsfonts}

\newcommand{\be}{\begin{equation}}
\newcommand{\ee}{\end{equation}}
\newcommand{\bea}{\begin{eqnarray}}
\newcommand{\eea}{\end{eqnarray}}

\newcommand{\gapp}{\mathrel{\raise.3ex\hbox{$>$}\mkern-14mu
              \lower0.6ex\hbox{$\sim$}}}
\newcommand{\lapp}{\mathrel{\raise.3ex\hbox{$<$}\mkern-14mu
              \lower0.6ex\hbox{$\sim$}}}

\begin{document}
\title{Gravity Waves Seeded by Turbulence and Magnetic Fields From a First Order Phase Transition With Non-Renormalizable Electroweak Vacua }
\author{Robert Poltis}
\affiliation{HEPCOS, Department of Physics, SUNY at Buffalo, Buffalo, NY 14260-1500}

\begin{abstract}

It is widely believed that the standard model is a low energy effective theory which may have higher dimensional non-renormalizable operators. The existence of these new operators can lead to interesting dynamics for the evolution of the universe, including the appearance of new vacuum states. If the universe today exists in a false vacuum, there will be a non-zero probability to tunnel to the true vacuum state of the universe. Should this transition occur elsewhere in the universe, bubbles of true vacuum will nucleate and expand outwards. Bubbles that nucleate in the hot dense plasma of the early universe will feel a friction from the plasma that acts against the expansion of the bubble, until the bubble eventually reaches a steady state expansion. Unlike many bubble formation scenarios where the bubble wall velocity rapidly approaches the speed of light, friction from the hot primordial plasma can cause the expanding bubble wall to reach a terminal velocity while gravity waves are free to propagate through the hot dense plasma at the speed of light. We analyze the effects of friction on the spectrum of gravity waves caused by bubble collisions. We find that a phase transition in a model with $\phi^6$ and $\phi^8$ operators that proceeds via a detonation in the hot plasma of the early universe is unlikely. In the case of a deflagration, the gravity wave spectrum is small and would likely require a post-LISA experiment such as the Big Bang Observer, but is in principle, observable.

\end{abstract}

\maketitle

\section{Introduction}

One of the most perplexing puzzles in all of science is the nature of the dark energy that is causing the accelerated expansion of the universe. The authors of \cite{Stojkovic:2007dw} pointed out that a late time phase transition could cause an accelerated expansion of the universe. In \cite{Greenwood:2008qp} a scalar field theory was presented which contained a standard model Higgs field, $\phi$, with a symmetric Higgs potential containing non-renormalizable operators up to $\phi^8$. In both works the late time accelerated expansion of the universe is explained as a result of the present day Higgs field residing in a metastable false vacuum. By adding non-renormalizable operators to the standard model Higgs potential, new vacuum states can form after the electroweak phase transition. The standard model electroweak phase transition may remain second order, afterwhich the universe ends up in a false vacuum situated above the true vacuum state (Fig.~(\ref{Phi8Potential})).

Due to the universe residing in a false vacuum state situated above a true vacuum, there exists a finite probability of bubbles of the true vacuum state spontaneously nucleating in the false vacuum permeated universe. If one of these bubbles is small, the tension of the bubble's surface will cause the bubble to collapse. Should the radius of one of these bubbles be greater than some critical value, however, the outward pressure pusing on the wall of the bubble will be greater than the tension of the bubble wall itself. In this case the bubble will grow in size as the wall of the bubble rapidly approaches the speed of light. If the nucleation rate per unit $3+1$-volume is sufficiently large, one might expect several bubbles of true vacuum to be formed in our current horizon. The collision of such bubbles could source gravity waves. The authors of \cite{Greenwood:2010wr} derived the characteristic frequency and gravity wave specturm expected from such a phase transition where the bubbles are nucleated via a detonation-i.e. the speed of the bubble wall is greater than the speed of sound in the medium the bubble wall is travelling through.

In this paper we consider the gravitational wave signature from turbulence and magnetic fields seeded by a bubble nucleation event for the $\phi^8$ model. In a detonation, the bubble wall expands faster than the speed of sound in the cosmological medium, while a deflagration is characterized by a bubble wall velocity less than the speed of sound in the cosmological medium. We find that turbulence of the cosmic fluid might be able to source gravity waves for a phase transition that proceeds via a deflagration, but the gravity wave signal is small. We also find that a first order phase transition that proceeds as a detonation, however, is unlikely in our model for the probable range of values of the Higgs mass as recently reported from the LHC.

The paper is organized as follows. Section II reviews the $\phi^8$ model proposed in \cite{Greenwood:2008qp}. In Section III we review the derivation of the relevant details of turbulence seeded by a first order phase transition for both detonations and deflagrations. Section IV reviews the major details of the calculation of the gravity wave energy density spectrum from turbulence and magnetic fields in three approximations. Section V calculates the gravity wave energy density spectrum in the deflagration case, and it is shown that detonations from a first order phase transition are unlikely in our model. Conclusions are presented in Section VI.

\section{Electroweak Model}

The authors of \cite{Greenwood:2008qp} introduced a standard model Higgs potential that contained higher order non-renormalizable terms. The problem of non-renormalizability may be (temporarily) ignored as long as such higher order terms are suppressed by large enough masses. The potential considered is (up to an irrelevant constant)
\be
V\left(\phi\right)=\frac{\lambda_3}{16}\phi^8-\frac{\lambda_2}{8}\phi^6+\frac{\lambda_1}{4}\phi^4-\frac{\mu^2}{2}\phi^2.
\label{IDontThinkWellUseThisLabelAtAll}
\ee
This potential may than be rewritten (again, up to an irrelevant constant) as
\be
V\left(\phi\right)=\frac{\lambda_3}{16}\left(\phi^2-\phi_1^2\right)^2\left(\phi^2-\phi_2^2\right)^2-\epsilon_o\phi_1^3\phi_2^2\phi^2,
\label{V}
\ee
with the identification 
\be
\phi_1^2\equiv\frac{2}{\lambda_3\phi_2^2}\left(\lambda_1-\frac{\lambda_2^2}{4\lambda_3}\right)
\label{Phi1}
\ee
\be
\phi_2^2\equiv\frac{\lambda_2}{2\lambda_3}\left(1+\sqrt{3-\frac{8\lambda_1\lambda_3}{\lambda_2^2}}\right)
\label{Phi2}
\ee
\be
\epsilon_o\equiv\frac{\mu^2}{2\phi_1^3\phi_2^3}-\frac{\lambda_3}{8\lambda_1\lambda_2}\left(\phi_1^2+\phi_2^2\right).
\label{Epsilon}
\ee
The advantage of writing the potential as in (\ref{V}) is that the parameter $\epsilon_o$ acts as a controllable fine tuning. When $\epsilon_o=0$ the vacua $\pm\phi_1$ and $\pm\phi_2$ are degenerate. For positive $\epsilon_o$, the $\pm\phi_2$ vacuum is the true vacuum of our model. For values of $\epsilon_o$ that are slightly negative, $\pm\phi_1$ is the lower vacuum state of $\pm\phi_1$ and $\pm\phi_2$. As $\epsilon_o$ becomes more negative, the symmetry is restored and $\phi_{VEV}\rightarrow 0$. The difference in energy between the $\phi_1$ and $\phi_2$ vacuum states is
\be
\delta V=\epsilon_o\phi_1^3\phi_2^3\left(\phi_2^2-\phi_1^2\right)
\label{dV}
\ee
(assuming $\phi_2>\phi_1$).

\begin{figure}
\includegraphics[width=3.4in]{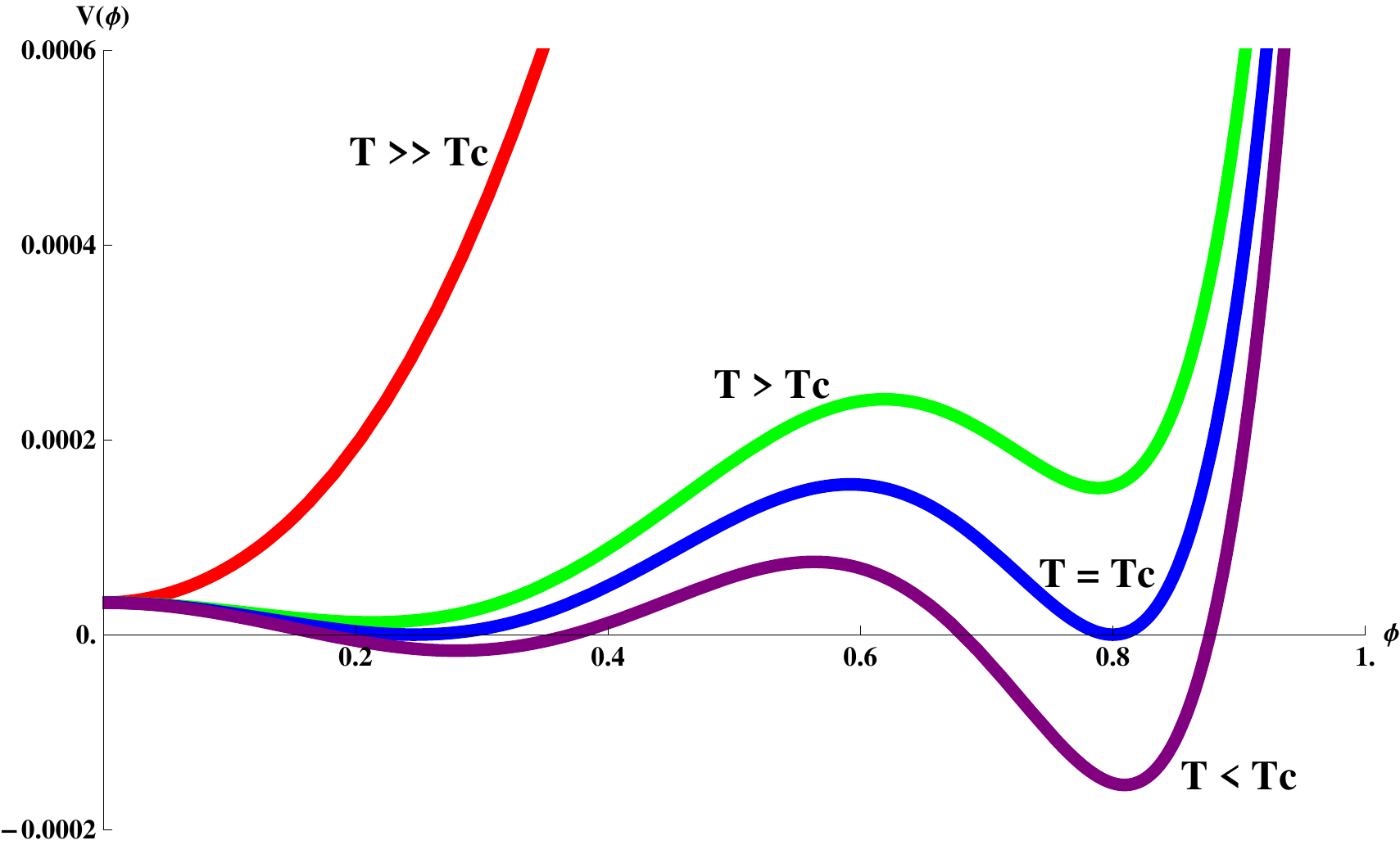}
\caption{The effective potential $V\left(\phi ,T\right)$ for four different temperatures. At very high temperatures (red line), the effective potential has a single minimum at $\phi=0$. As the temperature of the universe decreases, the electroweak phase transition may proceed as a second order phase transition with the Higgs field settling into the VEV at $\phi=\phi_1$. At some high temeperature (greater than the critical temperature), a second minima forms at $\phi=\phi_2$ (green line). At the critical temperature the vacua are degenerate (blue line). At temperatures below the critical temperature, the $\phi_2$ vacuum is the true vacuum of the potential (purple line). The field $\phi$ will be able to transition to the vacuum $\phi_2$ via a first order phase transition. The values used in this figure are $\phi_1=0.246$ TeV, $\phi_2=0.8$ TeV, $\lambda_3=0.35$ TeV$^{-4}$. The vaules of $\epsilon (T)$ shown are $-0.5$, $-0.025$, $0$, and $0.025$ from high to low temperature, respectively. The potential shown is symmetric about $\phi\rightarrow -\phi$.}
\label{Phi8Potential}
\end{figure}

Taking finite temperature effects into account amounts to adding a temperature-dependent thermal mass term of the form
\be
V(\phi ,T)=V(\phi ,0)+\frac{c}{2}T^2\phi^2.
\label{VofT}
\ee
The temperature-dependent effects may be absorbed into a parameter $\epsilon$ by setting
\be
\epsilon\left(T\right)=\epsilon_o-\frac{cT^2}{\phi_1^3\phi_2^3}.
\label{EpsilonT}
\ee
The constant $c$ is given in terms of $g$, $g^\prime$, and $y_t$ (the $SU(2)_L$, $U(1)_Y$ gauge couplings and top Yukawa coupling, respectively) by \cite{PRINT-78-0801 (SIMON-FRASER)}
\be
c=\frac{1}{16}\left(3g^2+g^{\prime 2}+4y_t^2+\frac{1}{32}\lambda_1\right).
\ee
From (\ref{EpsilonT}), a critical temperature may be defined where the vacua at $\pm\phi_1$ and $\pm\phi_2$ are degenerate:
\be
T_c^2=\frac{\epsilon_o\phi_1^3\phi_2^3}{c}.
\label{Tcrit}
\ee
The potential $V\left(\phi ,T\right)$ is shown in Fig.~(\ref{Phi8Potential}).
In this model, the field proceeds from a symmetric phase to a broken phase via a second order phase transition (from $\phi_{VEV}=0\rightarrow\phi_{VEV}=\pm\phi_1$). After the usual second order electroweak phase transition, the universe resides in the $\phi=\pm\phi_1$ vacuum state. As the temperature of the univese falls, a new mimima (located at $\phi=\pm\phi_2$) forms, and eventually becomes the true vacuum state of the universe (Purple line in Fig.~(\ref{Phi8Potential})). Because the two sides of the potential $V(\phi)$ are symmetric, from here on, we will only consider the positive half of the potential. Due to the barrier between the $\pm\phi_1$ and $\pm\phi_2$ minima, the field will eventually tunnel to the $\pm\phi_2$ ground state via a first order phase transition.

\section{Velocity Profile of the Plasma}
There are two mechanisms by which a first order phase transition may proceed: detonation or deflagration. A detonation is characterized by a supersonic wall velocity ($v_w>c_s$). The fluid in front of the expanding wall is taken to be at rest to an observer located at the center of the expanding bubble. The expanding bubble wall accelerates the fluid that it encounters, forming a rarefaction wave that trails the wall. A deflagration, on the other hand, is characterized by a subsonic wall velocity ($v_w<c_s$). In this case the fluid forms a shock front that preceeds the bubble wall, and comes to rest after the expanding bubble wall passes it.

As is standard, we take the speed of sound in the plasma to be $c_s^2=1/3$. Following the work of \cite{arXiv:1004.4187}, we relate the enthalpy ($w$), entropy density ($s$), energy density ($\rho$), temperature ($T$) and pressure ($p$) of the plamsa as
\be
w\equiv T\frac{\partial p}{\partial t}
\label{Enthalpy}
\ee
\be
s\equiv\frac{\partial p}{\partial t}
\label{EntropyDensity}
\ee
\be
\rho\equiv T\frac{\partial p}{\partial t}-p.
\label{EnergyDensity}
\ee
Due to the existence of a barrier between the $\phi_1$ and $\phi_2$ vacua, the universe will not immediately transition to the true $\phi_2$ vacuum when the temperature falls below the critical temperature. Instead, the universe remains in the metastable false vacuum ($\phi_1$) state for some length of time based on the decay rate of the false vacuum.

In the relativistic gas approximation, pressure and energy density in regions of the universe that reside in the $\phi_1$ vacuum are
\be
p_+=\frac{1}{3}a_+T_+^4-e
\label{p+}
\ee
\be
\rho_+=a_+T_+^4+e
\label{rho+}
\ee
where $e$ is the energy of the $\phi_1$ false vacuum. In the $\phi_2$ true vacuum state, the pressure and energy density are given by
\be
p_-=\frac{1}{3}a_-T_-^4
\label{p-}
\ee
\be
\rho_-=a_-T_-^4
\label{rho-}
\ee
with the vacuum energy density of the $\phi_2$ state set to zero and $+(-)$ denoting regions on the outside(inside) of the expanding bubble. The expressions given by (\ref{p+})-(\ref{rho-}) describe the 'bag equation of state'. Because there may be a different number of light degrees of freedom on either side of the bubble wall, $a_+$ will generically be different from $a_-$ (and more specifically $a_+>a_-$). For small values of $m_i/T$ (with i denoting the $i^{th}$ species), $a_\pm$ is given for the bag equation of state as
\be
a_\pm =\frac{\pi^2}{30}\sum\left[ N^b_i+\frac{7}{8}\lvert N^f_i\rvert\right]
\label{apm}
\ee
where the sum is over the $i$ light species and $N_i^{b(f)}$ is the number of internal degrees of freedom of the bosons (fermions). For particles with large $m_i/T$, the bag equation of state will still hold \cite{arXiv:1004.4187}.

The strength of a first order phase transition may be parameterized by the quantity $\alpha_+$ which is related to $a_+$ by
\be
\alpha_+\equiv\frac{e}{a_+T_+^4}.
\label{alpha+}
\ee

In the rest frame of an observer situated at the center of a detonation bubble, fluid outside the bubble wall (in the $\phi_1$ false vacuum) is at rest. In the rest frame of the bubble wall, fluid flows into the wall with a speed $v_+$, and leaves the bubble wall with a speed $v_-$ as the wall passes. For a bubble that is expanding via deflagration, however, the two velocities $v_+$ and $v_-$ are defined as the speed of the fluid in the wall frame at the edges of the shock front that preceeds the expanding bubble. In this case, the fluid behind the expanding bubble wall is at rest relative to a stationary observer located at the center of the expanding bubble after the bubble wall passes. In the wall rest frame these two speeds are related by
\begin{widetext}
\be
v_+=\frac{1}{1+\alpha_+}\left[\left(\frac{v_-}{2}+\frac{1}{6v_-}\right)\pm\sqrt{\left(\frac{v_-}{2}+\frac{1}{6v_-}\right)^2+\alpha_+^2+\frac{2}{3}\alpha_+-\frac{1}{3}}\right].
\label{v+}
\ee
\end{widetext}
with the $+$ appling to detonations and the $-$ sign appling to deflagrations.

\begin{figure*}
\centerline{
  \mbox{\includegraphics[width=2.90in]{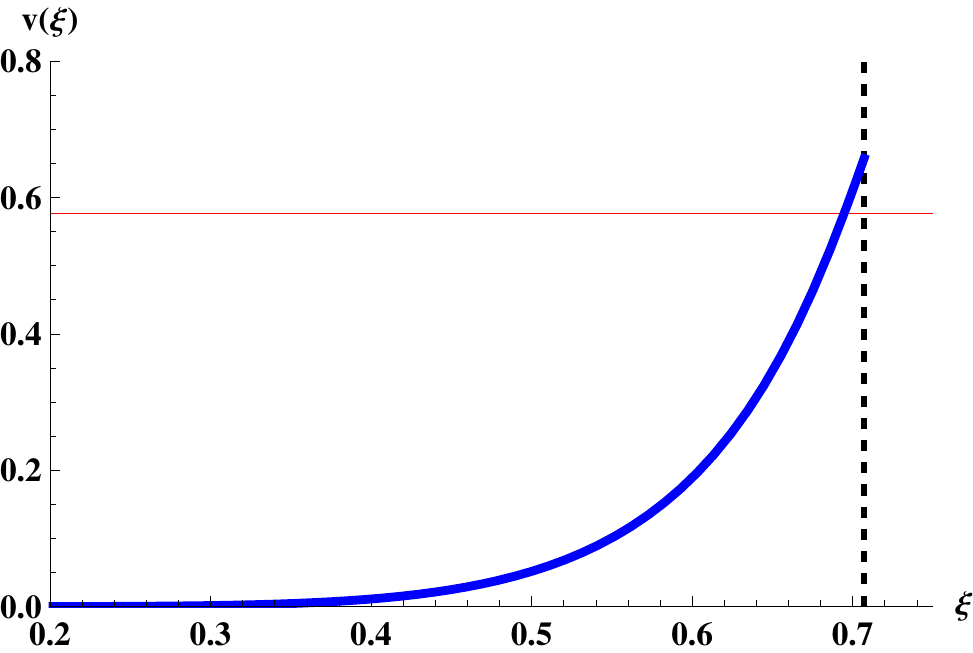}}
  \mbox{\includegraphics[width=2.90in]{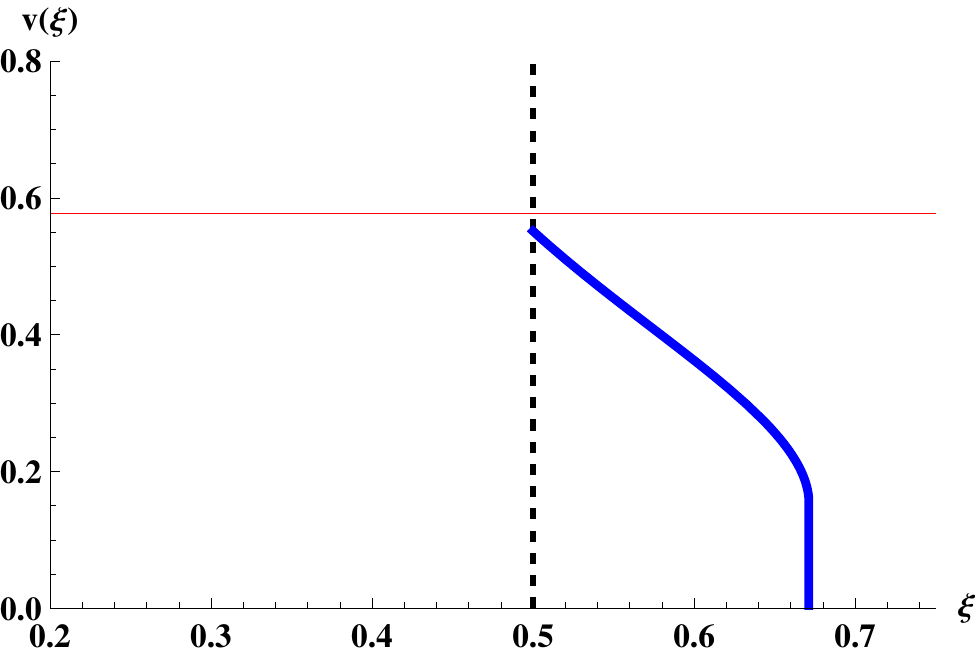}}
  }
\caption{The fluid velocity profile from (\ref{vxi}) of a detonation (left) and deflagration (right). The vertical dashed line is the location of the bubble wall which separates the $\phi_2$ vacuum on the left from the $\phi_1$ vacuum on the right. The red horizontal line is the speed of sound in the plasma ($c_s=1/\sqrt{3}$). The plots shown were calculated using $\alpha_+=0.02$ and $v_+=1/\sqrt{2}$ for the detonation case and $\alpha_+=0.0001$ and $v_+=0.55$ for the deflagration case.} 
\label{VelocityProfiles}
\end{figure*}

While (\ref{v+}) describes the fluid velocity in the immediate vicinity of the bubble wall and ends of a shock front, we will also want to know the full velocity profile of the fluid, as the calculation of gravity waves will be most strongly influenced by the largest fluid velocities in the plasma. As previously mentioned, a detonation bubble wall moving at supersonic speeds hits the stationary plasma and forms a rarefaction wave that trails behind the bubble wall. Deflagrations, on the other hand, cause the fluid to develop a shock front that travels in front of the expanding bubble. The authors of \cite{arXiv:1004.4187} derive an expression for the fluid velocity profile around an expanding bubble. The quantity $\xi$ is the velocity of a point in the wave profile given by $\xi\equiv r/t$, where $r$ is the distance from the bubble center and $t$ is the time elapsed since bubble nucleation. A particle at a point described by $\xi$ will move with a velocity $v(\xi )$ given by
\be
2\frac{v}{\xi}=\gamma^2\left(1-v\xi\right)\left(\frac{\mu^2}{c_s^2}-1\right)\frac{\partial v}{\partial \xi}
\label{vxi}
\ee
with $\mu$ the lorentz transformed fluid velocity $\mu(\xi ,v)=(\xi-v)/(1-\xi v)$. The fluid velocity profile of a detonation and deflagration are shown in Fig.~(\ref{VelocityProfiles}). 

Detonations and deflagrations may be categorized as weak, Jouguet, or strong based on the speed of the plasma at $v_-$, as shown in the table below.
\begin{center}
\begin{tabular}{|l|c|c|}
  \hline
    & Detonation & Deflagration \\
    \hline
  Weak & $v_->c_s$ & $v_-<c_s$ \\
  Jouguet & $v_-=c_s$ & $v_-=c_s$ \\
  Strong & $v_-<c_s$ & $v_->c_s$ \\
  \hline
\end{tabular}
\end{center}

Notice that while weak detonations are characterized by $v_->c_s$, weak deflagrations are characterized by $v_-<c_s$ (and similar for the strong case). Detonations and deflagrations may be either weak or Jouguet. Strong detonations and deflagrations are forbidden, and Jouguet phase transitions are also unlikely \cite{arXiv:0904.1753}-\cite{arXiv:0711.2593}. While not forbidden, \cite{KurkiSuonio:1995pp} found that in  numerical simulations expanding bubbles in cosmological phase transitions would encounter a discontinuous jump from weak deflagrations to weak detonations (essentially bypassing a Jouguet phase transition).

As bubbles of true vacuum nucleate, they will expand due to a difference in free energy on either side of the bubble wall. This driving force, however, is opposed by friction caused by deviations of particle distributions away from equilibrium in the surrounding plasma. Eventually an equilibrium is reached, whereafter the bubble continues to expand at a terminal velocity. The authors of \cite{arXiv:1004.4187} derive in a model-independent way the relationship between the bubble nucleation parameters and a friction parameter $\eta$:
\be
\alpha_+-\frac{1}{3}\left(1-\frac{a_-}{a_+}\right)=\eta\frac{\alpha_+}{\alpha_N}\left< v\right>,
\label{Friction}
\ee
where $\left< v\right>$ is the fluid velocity average across the bubble wall in the wall frame which may be approximated as
\be
\left< v\right>\equiv\frac{\int dz v\left(\partial_z\phi\right)^2}{\int dz\left(\partial_z\phi\right)^2}\simeq\frac{1}{2}\left(v_++v_-\right),
\label{VAvg}
\ee
$\alpha_N$ is given at the nucleation values by
\be
\alpha_N=\frac{e}{a_NT_N^4},
\label{AlphaN}
\ee
and the friction parameter $\eta$ is
\be
\eta\sim\frac{\hat{\eta}}{10a_+}\frac{1}{T_NL_{wall}}\left(\frac{\phi_N}{T_N}\right)^4
\label{Eta}
\ee
with $L_{wall}$ the bubble wall thickness.  The coefficient $\hat{\eta}$ was calculated to be $\approx 3$ in the SM \cite{hep-ph/9506475} and $\lesssim 100$ in the MSSM \cite{hep-ph/0002050}.

\section{Gravity Wave Spectrum}
Large bulk motions of fluid will source gravity waves which travel uninhibited from the time of their creation to present, much like the surface of last scattering has been able to travel through the universe unimpeded from its creation $\sim 300,000$ after the big bang to present day. The authors of \cite{arXiv:0909.0622} derive the gravity wave energy density spectrum of a long lasting source \footnote{A long-lasting source may be active for several Hubble times, as opposed to a short-duration source as in thin wall bubble collisions from detonations first studied by \cite{FERMILAB-PUB-91-333-A-REV}, \cite{FERMILAB-PUB-91-323-A}, and many works thereafter.} (such as magneto-hydrodynamic turbulence). Below we review the important results.

Working in conformal coordinates, peturbations of a flat FRW metric may be written as
\be
ds^2=a^2(t)\left[-dt^2+(\delta_{ij}+2h_{ij}dx^idx^j)\right].
\ee
Gravity waves are sourced by the transverse traceless part of the stress-energy tensor
\be
T_{ij}^{TT}\left( k,t\right)=\left(\rho +p\right)\Pi_{ij}\left( k,t\right)
\ee
where $\Pi_{ij}$ is defined through the projection tensor
\begin{eqnarray}
\Pi_{ij}\left({\bf k},t\right) = &&\!\!\!\! \left( P_{il}P_{jm}-1/2P_{ij}P_{lm}\right) T_{lm}\left({\bf k},t\right) \nonumber \\ && P_{ij}=\delta_{ij}-k_ik_j
\end{eqnarray}
In terms of the dimensionless variable $x_{1(2)}=kt_{1(2)}$, the gravity wave energy density spectrum of a (long lasting) source active from $t_{in}$ to $t_{fin}$ redshifted to today is
\begin{widetext}
\begin{eqnarray}
\left. h_o^2\frac{d\Omega_{GW}}{d\left( \mathrm{log}k\right)}\right|_o=\frac{4h_o^2\Omega_{rad,o}}{3\pi^2}\left(\frac{g_o}{g_{fin}}\right)^{1/3}k^3\int_{x_{in}}^{x_{fin}}\frac{dx_1}{x_1}\int_{x_{in}}^{x_{fin}}\frac{dx_2}{x_2}\mathrm{cos}\left( x_2-x_1\right)\Pi(k,x_1,x_2)
\label{GWSpectrumEquation}
\end{eqnarray}
\end{widetext}
where $g$ is the number of relativistic degrees of freedom and subscript $_o$ refers to the value of a quantity today. Thus, the calculation of the gravity wave energy density spectrum requires knowledge of the unequal time correlator of the tensor type anisotropic stress $\Pi(k,x_1,x_2)$.

The authors of \cite{arXiv:0909.0622} give three approximations to the GW energy density power spectrum: a coherent approximation, an incoherent approximation, and a top hat approximation.  The gravity waves source may be turned on or off via a funtion $f(t)$ of the form
\be
\Pi\left( k,t,t\right)=\left(\frac{\Omega_{s}}{\Omega_{rad}}\right)^2L^3\mathcal{S}\left( K\right)f^2\left( t\right)
\ee
with $K$ a dimensionless wavenumber related to the characteristic scale $L$ via $K=Lk/2\pi$. $\mathcal{S}(K)$ describes the scale dependence of the source and $\Omega_{s}$ is normalized to the present critical density. 

In what follows a subscript 's' will designate both a turbulent source (v) and a magnetic field source (b). The gravity wave energy density will be most strongly affected by the wavenumbers ($K_\ast =kL_\ast /2\pi$) corresponding to stirring on the largest scales $L_\ast$. The fluid on these scales has a characteristic velocity $v_L$. The gravity wave energy density spectrum is than given in terms of the wavenumber of the dimensionless time variable  $y=(t-t_{in})/\tau_L$, where $\tau_L$ is the typical turnover time of the largest turbulent eddies related to the time when turbulence becomes fully developed ($t_\ast$) by
\be
\frac{t_{in}}{\tau_L}=\frac{t_\ast}{\tau_L}-1,
\ee
and $\mathcal{C}_s$, a normalization constant of the turbulent velocity/magnetic power spectrum. At the time of the phase transition, the ratio of energy density of the source to the radiation energy density is ($\Omega_{s\ast}/\Omega_{rad\ast}$) and $\gamma$ is a constant with the value $\gamma =2/7$.

In the coherent approximation the source of gravity waves is correlated at all times, so that we may write
\be
\Pi\left( K,t_1,t_2\right)=\sqrt{\Pi\left( K,t_1,t_1\right)}\sqrt{\Pi\left( K,t_2,t_2\right)}.
\ee
In this case the coherent approximation gives the gravity wave energy density spectrum as
\begin{widetext}
\begin{eqnarray}
h_o^2\left.\frac{d\Omega_{GW}}{d\left( \mathrm{log}k\right)}\right|_o\!\!\!\!&&
=12\left(2\pi\right)^2\mathcal{C}_s^2\Omega_{rad,o}h_o^2\left(\frac{g_o}{g_{fin}}\right)^{\frac{1}{3}}\left(\frac{\Omega_{S\ast}}{\Omega_{rad\ast}}\right)^2K_\ast^3 \nonumber \\ && \times\left\{\left[\int^1_0dy\frac{y^{3\gamma /2+1}}{y+\frac{t_{in}}{\tau_L}}\sqrt{\mathcal{I}_s\left(K_\ast ,y,y\right)}\mathrm{cos}\left(\frac{\pi K_\ast}{v_L}y\right) +\int^{y_{fin}}_1dy\frac{y^{-7\gamma /2}}{y+\frac{t_{in}}{\tau_L}}\sqrt{\mathcal{I}_s\left(K_\ast ,y,y\right)}\mathrm{cos}\left(\frac{\pi K_\ast}{v_L}y\right)\right]^2\right. \nonumber \\ && +\left.\left[\int^1_0dy\frac{y^{3\gamma /2+1}}{y+\frac{t_{in}}{\tau_L}}\sqrt{\mathcal{I}_s\left(K_\ast ,y,y\right)}\mathrm{sin}\left(\frac{\pi K_\ast}{v_L}y\right) +\int^{y_{fin}}_1dy\frac{y^{-7\gamma /2}}{y+\frac{t_{in}}{\tau_L}}\sqrt{\mathcal{I}_s\left(K_\ast ,y,y\right)}\mathrm{sin}\left(\frac{\pi K_\ast}{v_L}y\right)\right]^2\right\}
\label{CoherentGWSpectrum}
\end{eqnarray}
\end{widetext}

The quantity $\mathcal{I}_s$ is given in the following subsections. The incoherent approximation is based on the assumption that the source of gravity waves is only correlated for times $t_1\simeq t_2$, and
\be
\Pi\left( K,t_1,t_2\right)=\Pi\left( K,t_1,t_1\right)\delta\left( t_1-t_2\right)\Delta tf^2\left( t\right)
\ee
where $\Delta t$ is a very short characteristic timescale that the source is approximately coherent. The incoherent gravity wave energy density spectrum is thus
\begin{widetext}
\begin{eqnarray}
h_o^2\left.\frac{d\Omega_{GW}}{d\left( \mathrm{log}k\right)}\right|_o &&
=12\left(2\pi\right)^2\mathcal{C}_s^2\Omega_{rad,o}h_o^2\left(\frac{g_o}{g_{fin}}\right)^{\frac{1}{3}}\left(\frac{\Omega_{S\ast}}{\Omega_{rad\ast}}\right)^2K_\ast^3 \nonumber \\ && \times\left(\int^1_0dy\frac{y^{3\gamma +2}}{\left(y+\frac{t_{in}}{\tau_L}\right)^2}\mathcal{I}_s\left(K_\ast ,y,y\right)+\int^{y_{fin}}_1dy\frac{y^{-7\gamma}}{\left(y+\frac{t_{in}}{\tau_L}\right)^2}\mathcal{I}_s\left(K_\ast ,y,y\right)\right)
\label{IncoherentGWSpectrum}
\end{eqnarray}
\end{widetext}
In \cite{astro-ph/0611894}, an analogue of Kraichnan's nonlocal spectrum of isotropic hydrodynaic turbulence arises due to incompressible magneto-hydrodynaic turbulence becoming nonlocal. The top hat approximation from \cite{arXiv:0711.2593} is the most relevant approximation to gravity waves sourced by magneto-hydrodynamic turbulence, as it mimics Kraichnan decorrelation better than the other two approximations. The top hat approximation assumes that $\Pi\left( k,t_1,t_2\right)$ is correlated for $\left| t_1-t_2\right| <x_c/k$ and uncorrelated otherwise, with $x_c$ a parameter of order unity.
\begin{widetext}
\begin{eqnarray}
\Pi\left( K_\ast,y,z\right) &&
=\frac{9}{4}\pi\mathcal{C}_s^2\left[\left(\frac{\Omega_{s}}{\Omega_{rad}}\left( y\right)\right)^2L^3\left( y\right)\mathcal{I}_s\left(K_\ast ,y,y\right)\Theta\left( z-y\right)\Theta\left(\frac{v_Lx_c}{\pi K_\ast}-\left( z-y\right)\right)\right. \nonumber \\ && + \left.\left(\frac{\Omega_{s}}{\Omega_{rad}}\left( z\right)\right)^2L^3\left( z\right)\mathcal{I}_s\left(K_\ast ,z,z\right)\Theta\left( y-z\right)\Theta\left(\frac{v_Lx_c}{\pi K_\ast}-\left( y-z\right)\right)\right]
\end{eqnarray}
\end{widetext}
Choosing the Kraichnan model value of $x_c=1$, the energy density of the gravity wave spectrum in the top hat approximation is than
\begin{widetext}
\begin{eqnarray}
h_o^2\left.\frac{d\Omega_{GW}}{d\left( \mathrm{log}k\right)}\right|_o&&
=12\left(2\pi\right)^2\mathcal{C}_s^2\Omega_{rad,o}h_o^2\left(\frac{g_o}{g_{fin}}\right)^{\frac{1}{3}}\left(\frac{\Omega_{S\ast}}{\Omega_{rad\ast}}\right)^2K_\ast^3 \nonumber \\ && \times\left[\int^1_0dy\frac{y^{3\gamma+2}}{y+\frac{t_{in}}{\tau_L}}\mathcal{I}_s\left(K_\ast ,y,y\right)\int^{y_{top}}_y\frac{dz}{z+\frac{t_{in}}{\tau_L}}\mathrm{cos}\left(\frac{\pi K_\ast}{v_L}\left( z-y\right)\right)\right. \nonumber \\ && + \left.\int^{y_{fin}}_1dy\frac{y^{-7\gamma}}{y+\frac{t_{in}}{\tau_L}}\mathcal{I}_s\left(K_\ast ,y,y\right)\int^{y_{top}}_y\frac{dz}{z+\frac{t_{in}}{\tau_L}}\mathrm{cos}\left(\frac{\pi K_\ast}{v_L}\left( z-y\right)\right)\right]
\label{TopHatGWspectrum}
\end{eqnarray}
\end{widetext}
with $y_{top}=\mathrm{min}\left[ y_{fin},y+\frac{v_Lx_c}{\pi K_\ast}\right]$. 

\subsection{Gravity Wave Energy Density Spectrum from Turbulence}
In (\ref{CoherentGWSpectrum}), (\ref{IncoherentGWSpectrum}) and (\ref{TopHatGWspectrum}) $\mathcal{I}_s\left(K_\ast ,y,y\right)$ is given by the numerical fit from \cite{arXiv:0909.0622} for turbulence by
\be
\mathcal{I}_v\left(K_\ast ,y,y\right)\simeq 0.098\left[ 1+\left(\frac{K_\ast y^\gamma}{4}\right)^{4/3}+\left(\frac{K_\ast y^\gamma}{3.3}\right)^{11/3}\right]^{-1}
\label{Iv}
\ee
The quantity $\Omega_{v\ast}/\Omega_{rad,\ast}$ is the ratio of the energy density from kinetic energy of turbulence to the radiation energy density at the time of the phase transition. Making the approximation that the turbulent fluid may be described by a single typical value $v$, with $v$ being the dominant contribution to the kinetic energy of the turbulent flow and writing $\left< v^2\right>\sim v^2$, the kinetic energy of the turbulent flow may be written as
\be
\rho_{kin}=\left( \rho +p\right)\frac{\left< v^2\right>}{2}
\ee
and the energy density of turbulence
\be
\left< v^2\right> =\frac{3}{2}\frac{\Omega_{v}}{\Omega_{rad}}.
\label{TurbEnergyDensity}
\ee

\subsection{Gravity Wave Energy Density Spectrum from Magnetic Fields}
In the magnetic case, $\mathcal{I}_s\left(K_\ast ,y,y\right)$ in (\ref{CoherentGWSpectrum}), (\ref{IncoherentGWSpectrum}) and (\ref{TopHatGWspectrum}) is best fit by
\be
\mathcal{I}_b\left(K_\ast ,y,y\right)\simeq 0.12\left[ 1+\left(\frac{K_\ast y^\gamma}{4}\right)^{4/3}+\left(\frac{K_\ast y^\gamma}{3.5}\right)^{7/2}\right]^{-1}
\label{Ib}
\ee
\cite{arXiv:0909.0622}. The ratio of the magnetic field energy density to radiation energy density $\Omega_{b\ast}/\Omega_{rad,\ast}$ is related through the normalized magneitc field vector $b_i$:
\be
\left< b^2\right> =\frac{3}{2}\frac{\Omega_{b}}{\Omega_{rad}}.
\label{MagEnergyDensity}
\ee
\be
b_i=\sqrt{\frac{3}{16\pi\rho_{rad}}}B_i
\ee

\section{Results}
\subsection{Detonation}
The tunnelling rate per unit 4-volume from false to true vacuum in the detonation case is 
\be
\Gamma =Ae^{-S_E}
\label{rate}
\ee
determined primarily by the Euclidean Action $S_E$:
\begin{eqnarray}
S_E && =2\pi^2\int_0^\infty r^3dr\left[\frac{1}{2}\left(\frac{d\phi}{dr}\right)^2+V_{eff}\left(\phi\right)\right] \nonumber \\ && =-2\pi^2\frac{R^4}{4}\delta V+2\pi^2 R^3\int_{\phi_1}^{\phi_2}d\phi\sqrt{2V\left(\phi ,T_c\right)}
\label{EuclideanAction}
\end{eqnarray}
where the first line of (\ref{EuclideanAction}) is true in general and the second line of (\ref{EuclideanAction}) is true in the thin wall approximation evaluated at $T_c$, where the bubble wall thickness is small compared to the bubble radius \cite{Kolb:1990vq}. The term $A$ in (\ref{rate}) is a dimensionful constant that depends on loop corrections of (\ref{IDontThinkWellUseThisLabelAtAll}), but because we are only interested in general behavior, we will concentrate on the behavior of the action $S_E$. We require the region of the universe that we reside in to remain in the $\phi_1$ false vacuum, and therefore we require $t_{Hubble}^4\Gamma <1$. Following \cite{Coleman:1977py}, the radius of a critical bubble at lowest order in $\epsilon$ is
\begin{eqnarray}
R\left( T=0\right) \!\! && \!\!\!\! =\frac{3\int^{\phi_2}_{\phi_1}d\phi\sqrt{2V\left(\phi ,T_c\right)}}{\delta V\left( T\right)} \nonumber \\ \!\! && \!\!\!\! = \frac{2}{5}\sqrt{\frac{\lambda_3}{8}}\frac{\left(\phi_2-\phi_1\right)^2\left(\phi_1^2+3\phi_1\phi_2+\phi_2^2\right)}{\epsilon_o\phi_1^3\phi_2^3\left(\phi_1+\phi_2\right)}.
\label{RCrit0}
\end{eqnarray}
From (\ref{RCrit0}) the Euclidean action is
\be
S_E=\frac{2\pi^2\lambda_3^2}{50625}\frac{\left(\phi_2-\phi_1\right)^9\left(\phi_1^2+3\phi_1\phi_2+\phi_2^2\right)^4}{\epsilon_o^3\left(\phi_1+\phi_2\right)^3\phi_1^9\phi_2^9}
\label{SET0}
\ee
Setting $t_{Hubble}\sim10^{17}$s, the probability that our region of the universe resides in the false vacuum is satisfied for $S_E>160$.

In the finite temperature case, the temperature dependent radius of a critical bubble is
\begin{eqnarray}
R\left( T\right) && \!\!\!\! =\frac{2\int^{\phi_2}_{\phi_1}d\phi\sqrt{2V\left(\phi ,T_c\right)}}{\delta V\left( T\right)} \nonumber \\ && \!\!\!\! = \frac{\sqrt{2\lambda_3}\left(\phi_2-\phi_1\right)^2\left(\phi_1^2+3\phi_1\phi_2+\phi_2^2\right)}{15\epsilon\left( T\right)\phi_1^3\phi_2^3\left(\phi_1+\phi_2\right)}.
\label{RCritT}
\end{eqnarray}

A first order phase transition of the model described by (\ref{IDontThinkWellUseThisLabelAtAll}) that proceeds as a detonation, however is unlikely. Due to the fact that the phase transition front of a detonation propagates faster than the speed of sound in the plasma, the nucleation temperature is the same as the temperature in the $\phi_1$ false vacuum: $T_N=T_+$; and similarly, $\alpha_N=\alpha_+$. Furthermore, we will assume the standard model value for $\hat{\eta}$, $\mu =0.125$TeV, the thickness of the wall is not very different from what was found in \cite{Greenwood:2008qp}: $L_{wall}\sim 40 TeV^{-1}$, and that the false vacuum is at $\phi_1=0.49$TeV. To find $a_\pm$ (\ref{apm}), we approximate $N_+\sim 100$, which gives $a_+\sim 33$. To determine $a_-$, we assume that in the $\phi_1$ vacuum all particles are relativistic while in the $\phi_2$ vacuum the Higgs, W${^\pm}$, Z$_0$ and top quark are nonrelativistic. The difference of internal degrees of freedom across the two vacua
\be
\Delta N=N_1-N_2=\left( N^b_1+\frac{7}{8}N^f_1\right)-\left( N^b_2+\frac{7}{8}N^f_2\right)
\label{DeltaN}
\ee
is $\Delta N\sim 15$. From this we estimate we find $a_-\sim 28$. 

In a detonation, a bubble of true vacuum is nucleated in the false vacuum, and therefore $\alpha_+=\alpha_N$, $a_+=a_N$ and $T_+=T_N$. Combining (\ref{Friction}) and (\ref{Eta}) gives
\be
v_++v_-=\frac{8800T_+^5}{\phi_2^4}\left(\alpha_+-\frac{5}{99}\right).
\label{vtot}
\ee
The strength of the phase transition in this case is given by
\be
\alpha_+=\frac{\delta V}{a_+T_+^4}
\label{alpha+ex}
\ee

A requirement for our model presented here is that our local region of the universe resides in the $\phi_1$ false vacuum. For this requirement to be realized, the $\phi_1$ false vacuum must remain a false vacuum as $T\rightarrow0$. Furthermore, the fluid velocities in a detonation phase transition must satisfy
\be
\sqrt{3}/2<v_++v_-<2.
\label{DetReq}
\ee
\begin{figure*}[hbtp]
\begin{center}
\includegraphics[width=7.0in]{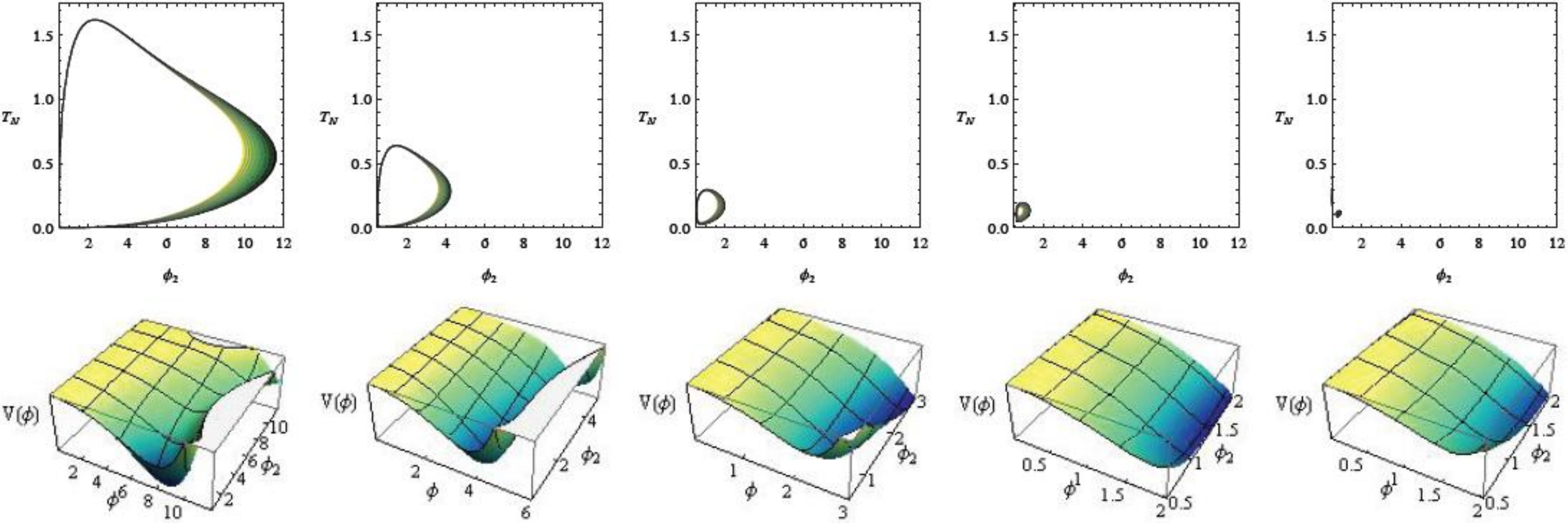}
\caption{Top Row: Values of $\phi_2$ and $T_N$ that satisfy the velocity requirement of a detonation solution ($2/\sqrt{3}<v_++v_-<2$). The true vacuum of the theory, $\phi_2$, is shown from $0.49$TeV to $12$TeV and the bubble nucleation temeperature runs from $0$ to $1.75$TeV.
\newline Bottom Row: The Potential $V(\phi)$ at $T=0$ (\ref{V}) for different values of the true vacuum $\phi_2$.
\newline Values of $\phi_2$ that satisfy the velocity requirement of a detonation solution (according to the top row) are unlikely because the potential $V(\phi)$ does not possess a false vacuum $\phi_1$ (that we must presently reside in) at $T=0$. In both rows, from left to right we set $\lambda_3=0.0001$, $0.001$, $0.005$, $0.010$ and $0.015$ respectively.}
\label{NoDetonSol}
\end{center}
\end{figure*}

The top row of Fig. (\ref{NoDetonSol}), shows values of $\phi_2$ and $T_+$ that satisfy (\ref{DetReq})) in a detonation for different choices of $\lambda_3$. However, the requirement that the false vacuum state $\phi_1$ still exists today (at $T=0$) cannot be satisfied for the small values of $\lambda_3$ (or similarly large values of $\phi_2$) that allow for both a false and true vacuum. For $\lambda_3>0.017$, there are no values of $\phi_2$ and $T_+$ that satisfy (\ref{vtot}), (\ref{DetReq}) and still allow the existence of a false vacuum at $\phi_1$ and true vacuum at $\phi_2$. In conclusion, combinations of $\phi_2$, $\lambda_3$, and $T_+$ that satisfy (\ref{DetReq}) do not appear to allow for the existence of a false vacuum state at zero temperature; and therefore detonation solutions that reach a terminal velocity appear unlikely in our model.

\subsection{Deflagration}
The deflagration model, in contrast to the detonation model, is not as simple. Because the bubble front is preceeded by a shock front, the temperature of the plasma immediately in front of the expanding bubble wall is increased, thus decreasing the energy separation of the $\phi_1$ and $\phi_2$ minima. We consider a model in which $\phi_1=0.49$TeV, $\phi_2=1.56$TeV, $\mu=0.125$TeV and $\lambda_3=0.2$TeV$^{-4}$.

In the deflagration case, the probability of a transition from a false to a true vacuum per unit 4-volume is governed by the expression \cite{arXiv:0904.1753} 
\be
\Gamma (T)\simeq A(T) e^{-S_E}
\label{DecayRate}
\ee
with
\be
A(T)=\left(\frac{S_3(T)}{2\pi T}\right)^{3/2},
\label{DecayRateCoefficient}
\ee
and $S_3$ the three dimensional instanton action. Near the critical temperature, $S_E$ can be approximated \cite{arXiv:0804.0391} by
\be
S_E\approx\frac{S_3(T)}{T}\approx\frac{16\pi\sigma^3T_c}{3\ell^2\left( T_c-T\right)^2}
\label{S3}
\ee
where $\ell$ is the difference in energy density between the two vacua at critical temperature and $\sigma$ is the bubble wall surface tension.

To calculate the bubble surface tension, we concentrate on a section of the bubble wall which is approximately planar lying perpendicular to the z direction. The stress tensor
\be
T_{\mu\nu}=\partial_\mu\phi\partial_\nu\phi -\mathcal{L}g_{\mu\nu}
\ee
can be found using the Bogomolnyi equation \cite{Vachaspati:2006zz} which gives the kink solution of the field that interpolates between the $\phi_1$ and $\phi_2$ vacua:
\be
\partial_z\phi\pm\sqrt{2V\left(\phi\right)}=0.
\label{Bogomolnyi}
\ee
The tension in the domain wall is than
\be
\sigma =\frac{\lambda_3}{8}\phi_1^4\phi_2^4.
\label{tension}
\ee
Although the vacuum energy of the two vacua is degenerate at critical temperature, different number of relativistic species in the two vacua will lead to a difference in the energy density across the bubble wall. Setting
\be
\ell=\left( a_+-a_-\right)T_c^4,
\label{ell}
\ee
and again taking $a_+\sim 33$ and $a_-\sim 28$ gives $\ell\sim 121$MeV.
Combining (\ref{apm}), (\ref{DeltaN}), (\ref{S3}), (\ref{tension}) and (\ref{ell}) we find
\be
\frac{S_3(T)}{T}\approx\frac{\lambda_3^3\phi_1^{12}\phi_2^{12}}{24\pi^3T_c^7\left( T_c-T\right)^2}.
\label{S3Calc}
\ee
As before, we will require $t^4\Gamma <1$, but this time we will look at a time around the critical temperature. We want to ensure that our local region of the universe did not yet go through a phase transition. We will take $t\sim t_{TeV}\sim 10^{-12}s$, which gives $t^4_{TeV}\Gamma\ll 1$.

The number of degrees of freedom on either side of the shock front that preceeds a deflagration bubble wall is the same because both sides of the shock front remain in the same vacuum state. Therefore, in the case of a deflagration, we may set $a_+=a_-$. Using (\ref{Friction}) and (\ref{Eta}), and assuming that the nucleation temperature is close to the critical temperature (here we use $T_N=80.3$GeV), the strength of the phase transition is $\alpha_+=0.331$ and the total fluid velocity across the shock front is $v_++v_-=0.00164$. From (\ref{v+}) this corresponds to a maximum fluid velocity of $v_+\approx 0.00164$. 

\begin{figure*}[hbtp]
\begin{center}
\mbox{
	\leavevmode
	\subfigure
	{ \label{deflagvcohsub}
	  \includegraphics[width=3.4in]{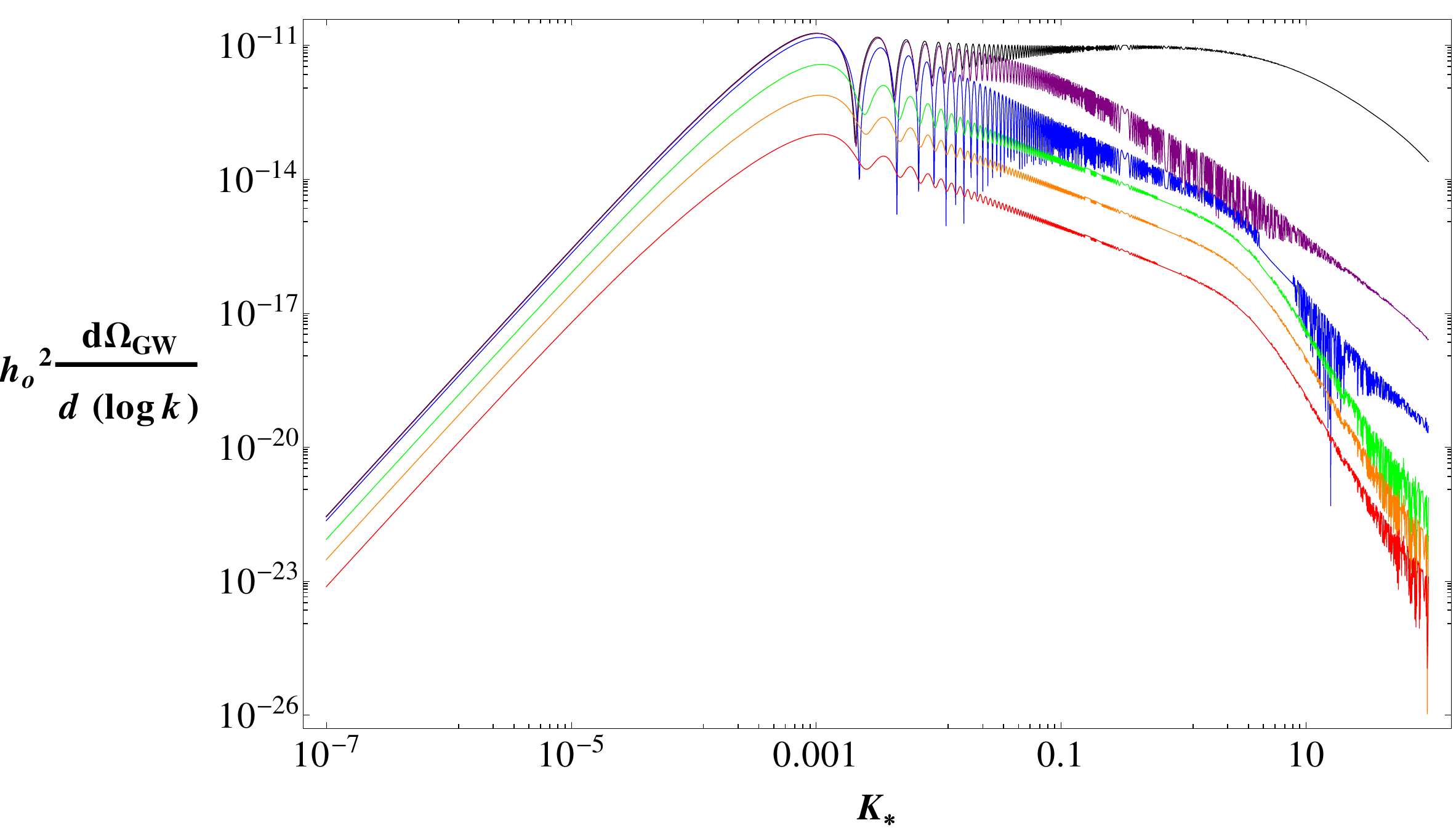} }

	\leavevmode
	\subfigure
	{ \label{deflagbcohsub}
	  \includegraphics[width=3.4in]{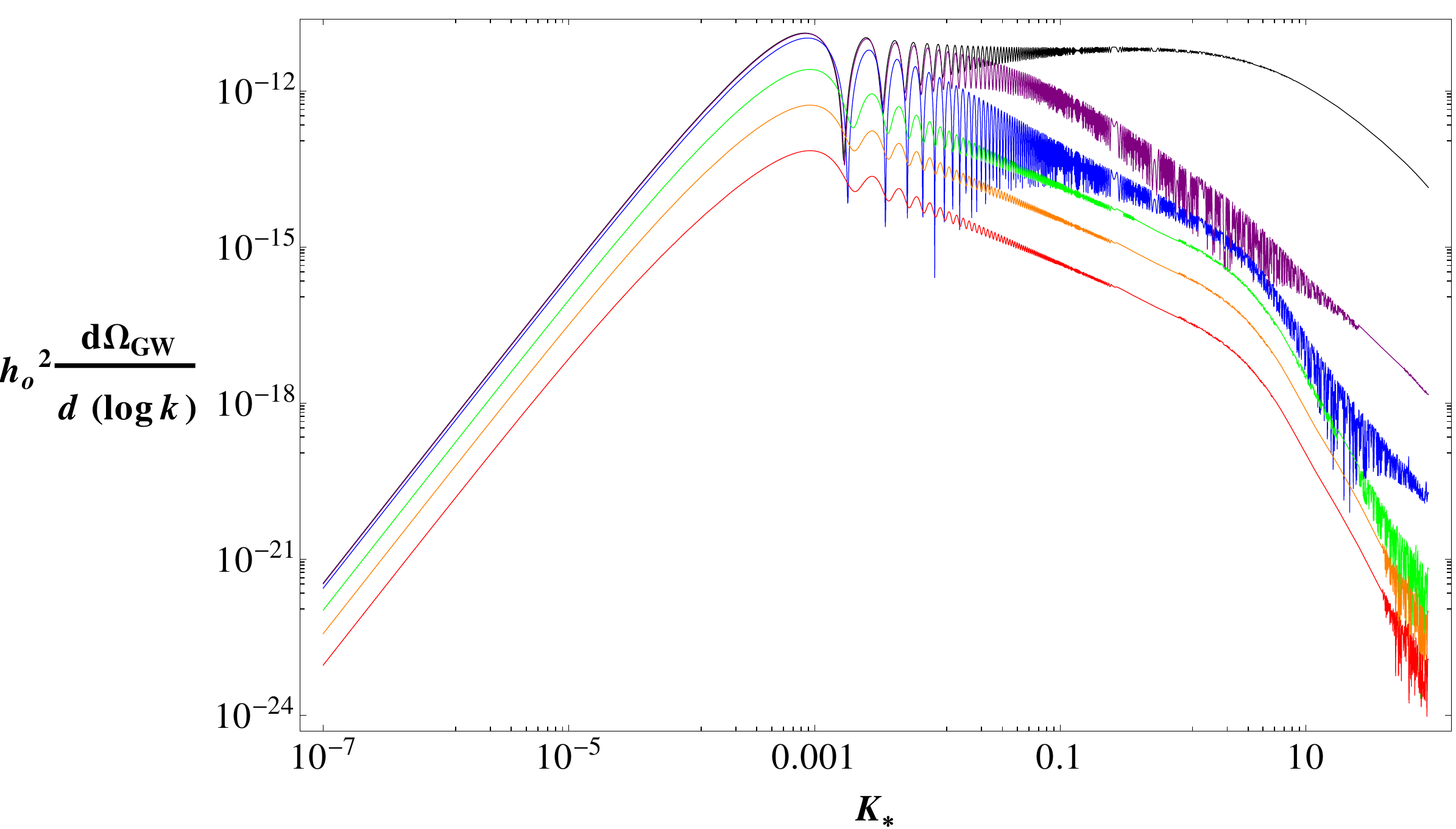} }
	}  

\mbox{
	\leavevmode
	\subfigure
	{ \label{deflagvincohsub}
	  \includegraphics[width=3.4in]{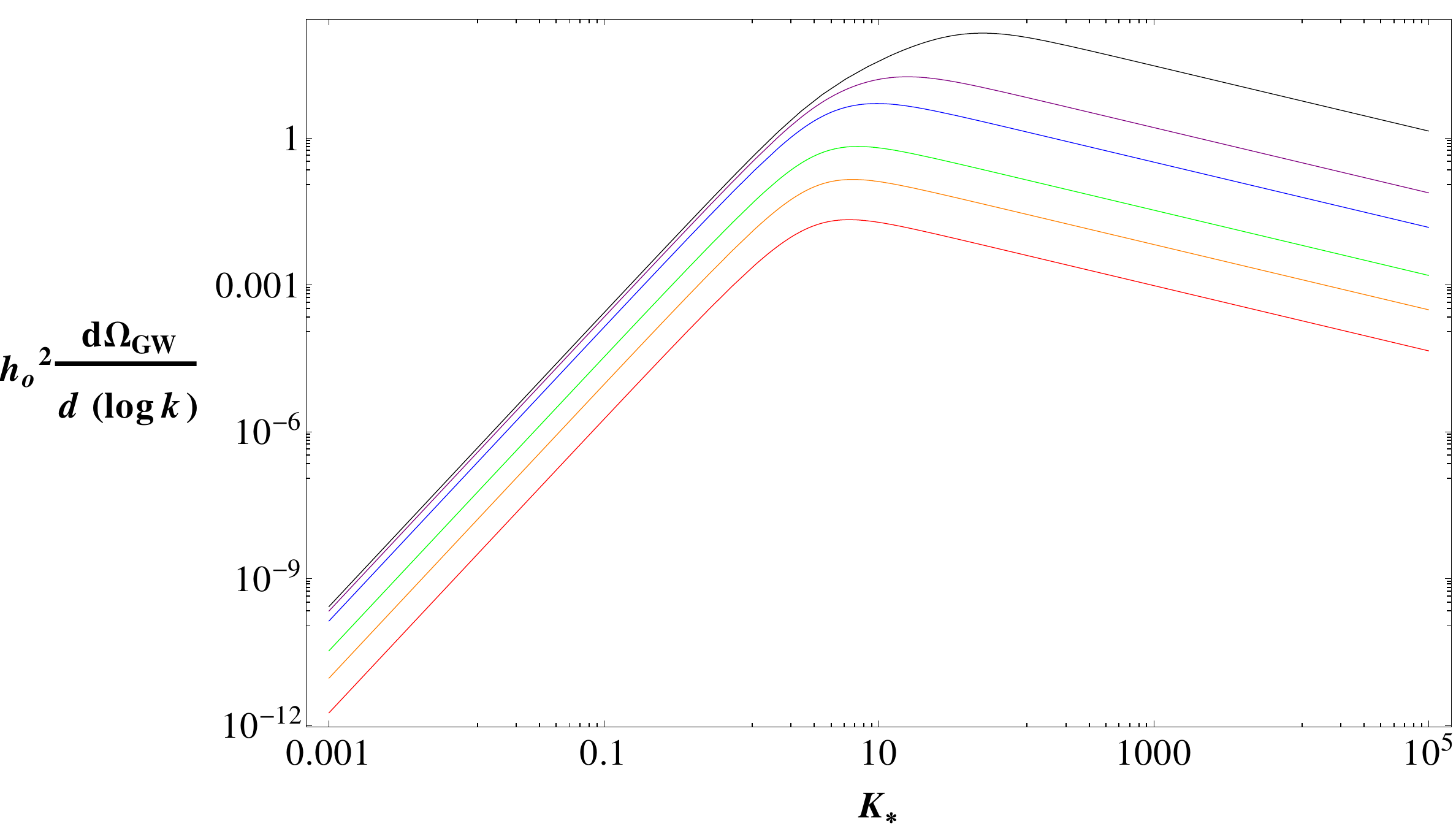} }

	\leavevmode
	\subfigure
	{ \label{deflagbincohsub}
	  \includegraphics[width=3.4in]{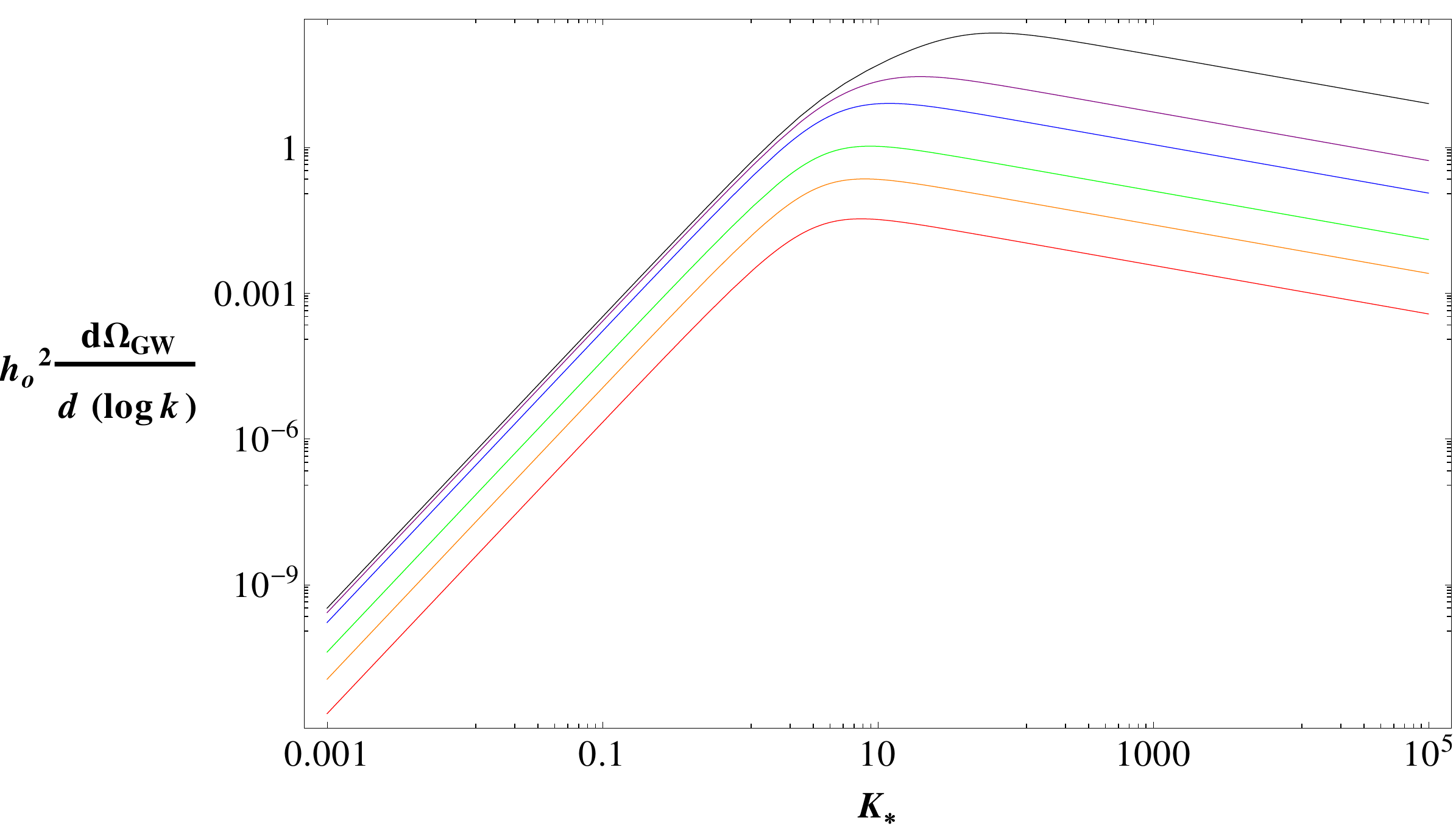} }
	}  

\mbox{
	\leavevmode
	\subfigure
	{ \label{deflagvtophatsub}
	  \includegraphics[width=3.4in]{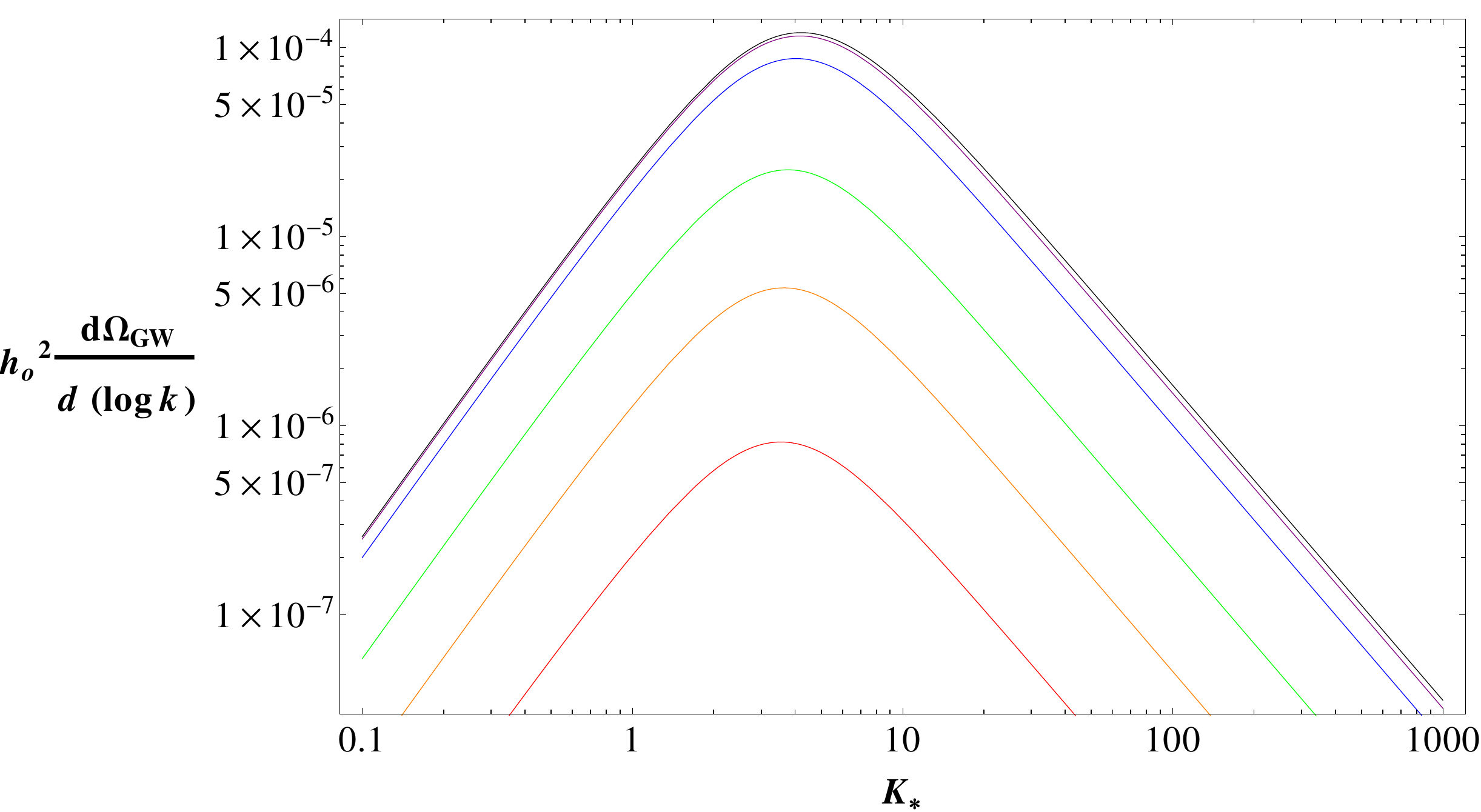} }

	\leavevmode
	\subfigure
	{ \label{deflagbtophatsub}
	  \includegraphics[width=3.4in]{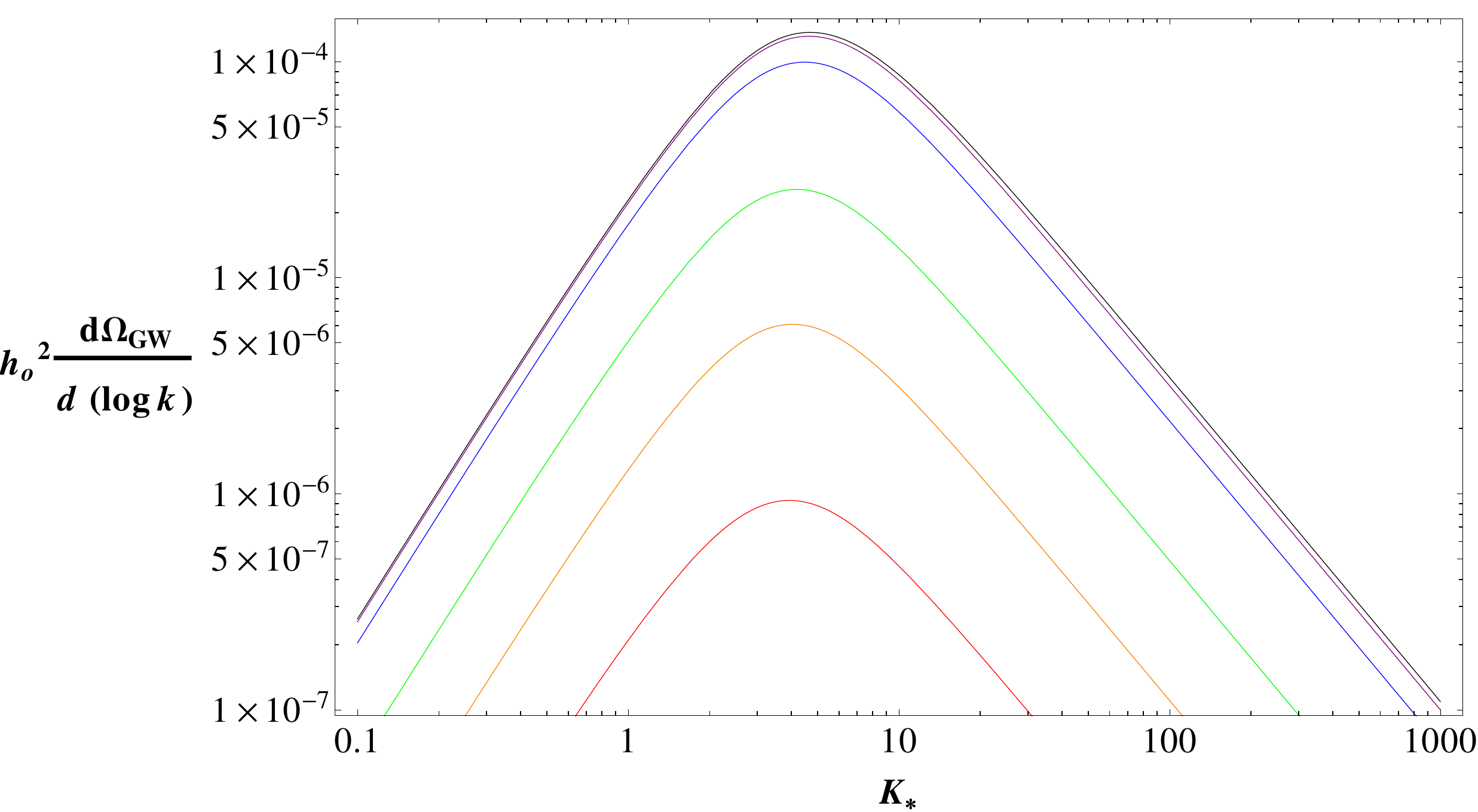} }
	}
	
\caption{The gravity wave energy density spectrum $\left( h_o^2\left.\frac{d\Omega_{GW}}{d\left( \mathrm{log}k\right)}\right|_o\right)$ of the deflagration phase transition described above evaluated today related to the dimensionless wavenumber $K_\ast$ for the deflagration phase transition described above. The top row is the gravity wave energy density spectrum sourced by turbulence (left) and magnetic fields (right) in the coherent approximation (\ref{CoherentGWSpectrum}). The middle row is that same as above, but evaluated in the incoherent approximation (\ref{IncoherentGWSpectrum}); and bottom row is evaluated in the top hat approximation (\ref{TopHatGWspectrum}). The gravity wave energy density spectrum is given in terms of  $12\left(2\pi\right)^2\mathcal{C}_s^2\Omega_{rad,o}h_o^2\left(\frac{g_o}{g_{fin}}\right)^{\frac{1}{3}}\left(\frac{\Omega_{S\ast}}{\Omega_{rad\ast}}\right)$, which is $\approx 2.23\times 10^{-17}$ for turbulence and $\approx 5.03\times 10^{-18}$ for magnetic fields. The different colors represent different values for $\frac{t_{in}}{\tau_L}$, with $\frac{t_{in}}{\tau_L}=10^{-4}$, $10^{-2}$, $10^{-1}$, $1$, $10^{1/2}$ and $10^{2}$ for black, purple, blue, green, orange, and red (respectively).}
\label{Deflag}

\end{center}
\end{figure*}

In the analysis we set $h_o^2\Omega_{rad,o}\approx 4.2\times 10^{-5}$ and $g_o\approx g_{fin}$. The characteristic velocity of the largest stirring scale is than $v_L\simeq 0.00114$. The normalization constant of the turbulent power spectrum is (see section 3.2 of \cite{arXiv:0711.2593}) $\mathcal{C}_v\approx 0.385$ and $\Omega_{v\ast}/\Omega_{rad\ast}=8.69\times 10^{-7}$ from (\ref{TurbEnergyDensity}). For the case of magnetic fields we assume equipartition between the turbulent and magnetic energy densities ($\left< v^2\right>\simeq\left< b^2\right>$). The normalization constant of the magnetic power spectrum is $\mathcal{C}_b \approx 0.0266$, $v_L\simeq 0.00095$ and $\Omega_{b\ast}/\Omega_{rad\ast}=5.99\times 10^{-7}$. In fig. (\ref{Deflag}), the gravity wave energy density spectrum are shown for both the turbulent case and magnetic case in the coherent, incoherent, and top hat approximations.

\section{Conclusion}

We have considered a modified Higgs potential wherein the usual second order electroweak phase transition in the early universe is followed by a first order phase transition sometime later. In many models of bubble nucleation events, the bubble wall velocity rapidly approaches the speed of light, thereby leaving no time to observe the gravitational wave signature of such a phase transition before the bubble wall passes through the Earth. By requiring the phase transition to have occurred in the early universe, the hot plasma may act as a frictional force on an expanding bubble of true vacuum causing the bubble wall to reach a terminal velocity while gravity waves are free to propagate out at the speed of light from the time of their formation. This would allow the gravity waves to be observable in gravity wave detectors long before the bubble reaches the detector.

Through assuming that our local region of the universe still resides in a false vacuum state, we have demonstrated that a first order phase transition that proceeded via a detonation within our past horizon is unlikely. The universe may still have tunnelled to the true vacuum somewhere within our causal horizon, but it will likely have done so as a deflagration rather than as a detonation. In our example the bubble wall reachs a very subsonic steady-state expansional velocity, which in turn causes the amplitude of the gravity wave energy density spectrum to be small. 

In our model, the gravity wave energy density spectrum in the incoherent approximation gives a maximum $h_o^2\left.\frac{d\Omega_{GW}}{d\left( \mathrm{log}k\right)}\right|_o\sim 10^{-17}$ and $\sim 10^{-18}$ in the incoherent approximation for gravity waves sourced by turbulence of the primordial plasma and magnetic fields, respectively. In the top hat approximation, $h_o^2\left.\frac{d\Omega_{GW}}{d\left( \mathrm{log}k\right)}\right|_o\sim 10^{-21}$ and $\sim 10^{-22}$ from turbulence and magnetic fields, respectively. This is well below the expected detector sensitivity of LISA, and would require a more ambitous experiment such as the Big Bang Observer \cite{Delaunay:2006ws}. 

Of the three approximations to the gravity wave energy density spectrum presented in Fig.~(\ref{Deflag}), the coherent approximation stands out in that it predicts a much smaller amplitude ($h_o^2\left.\frac{d\Omega_{GW}}{d\left( \mathrm{log}k\right)}\right|_o\sim 10^{-28}$ and $\sim 10^{-29}$ for turbulence and magnetic fields) compared to the incoherent and top hat approximations. The coherent approximation, however, is likely the least suited to calculate $h_o^2\left.\frac{d\Omega_{GW}}{d\left( \mathrm{log}k\right)}\right|_o$ from turbulence and magnetic fields. In the case of bubble wall collisions (from detonations), a single collision event may be considered coherent. A large number of bubble collisions is than just a sum of many coherent events, each of which source gravity waves well described by the coherent approximation, even if the events are not themselves correlated.

Gravity waves sourced by turbulence and magnetic fields, on the other hand, are not well described in the coherent approximation. Correlations are expected to decay over a time comparable to the spacial extent of the source, which in our case is small due to the bubble reaching a low terminal velocity. The bubble will continue to expand, but the low velocity (which implies smaller size and therefore less correlation) will tend to favor the incoherent approximation in this case. The top hat approximation in this regard can be considered a variation of the incoherent approximation, as the fluid dynamics are correlated at some, but not all times during the sourcing of gravitational waves.

\begin{acknowledgments}
R.P. would like to thank Eric Greenwood, Evan Halstead and Dejan Stojkovic for helpful discussion, and the physics department at Case Western Reserve University for their hospitality.
\end{acknowledgments}

\end{document}